\documentstyle[aps,prl,twocolumn]{revtex}

\draft

\begin{document}

\title{ Phase diagram of quantized vortices in a trapped 
Bose-Einstein condensed gas}
\author{ S. Stringari}
 \address{Dipartimento  di Fisica, Universit\`{a} di Trento,}
\address{and Istituto Nazionale di Fisica della Materia,
I-3850 Povo, Italy}
 \date{ 19 December 1998}

 \maketitle

 \begin{abstract}
 We investigate the thermodynamic stability of quantized vortices  in a dilute
Bose
gas confined by a rotating harmonic trap at finite temperature.
Interatomic forces play a crucial role in characterizing the resulting
phase diagram, especially in the large $N$ Thomas-Fermi
regime. We show that the critical temperature for the creation of stable
vortices exhibits a maximum as a function of the frequency of the rotating
trap and that
the corresponding transition is associated with a discontinuity in the
number of atoms in the condensate. Possible strategies
for approaching the vortical region are discussed.
 \end{abstract}


\narrowtext

The occurrence of quantized vortices in superfluids has been the object
in the past of
fundamental theoretical
and experimental  work
\cite{onsager,feynman,vinen,packard,donnelly}.
After the experimental realization of
Bose-Einstein condensation in trapped alkali gases
  \cite{bec}, the structure of vortices in these novel many-body sytems 
has soon attracted
the attention of theorists \cite{baym,dalfovo,burnett} (for a recent review
on various theoretical aspects of Bose-Einstein condensation in
trapped gases, see \cite{rmp}).
Special attention has been devoted, in particular, to the
calculation of the critical
frequency needed to create a stable vortex in the rotating frame.  
This quantity
is
of crucial importance in view of the experimental possibility of creating
vortices by rotation of the confining trap.
In systems interacting with repulsive forces
the critical frequency turns out to be smaller than the oscillator
frequency characterizing the confining harmonic potential and to decrease
smoothly with the number of  atoms in the condensate.
The purpose of this work is to discuss the behavior of the critical
frequency at finite temperature, pointing out the crucial role played
by the interactions, especially in the large N, Thomas-Fermi regime,
where the system behaves like
a Bogoliubov gas.
We will show that, due to the  harmonic nature of the confinement,
the phase diagram for the vortical configurations of these dilute Bose gases
exhibits new interesting  features which are absent in dense superfluids.

Let us consider a trap rotating with frequency $\omega$ along the $z$-axis.
The equilibrium configuration of the system in the rotating frame is  obtained
by minimizing the free energy $F=E-TS$, using the  Hamiltonian
$H(\omega) = H - \omega J_z$
where $H$ and $J_z$ are the Hamiltonian and the
angular momentum (third component) of the many-body system,
evaluated in the laboratory frame.
A vortical configuration is thermodynamically
stable if the trap rotates with frequency $\omega$
larger than the critical frequency
\begin{equation}
\omega_c = {F_{v} - F_0 \over \langle J_z\rangle_{v} - \langle J_z
\rangle_0} \, ,
\label{omegacF}
\end{equation}
fixed by the difference between the free energy
of the configurations with and without the vortex  and by the corresponding
difference of the angular momentum.
 The stability
criteria for vortices, including the possible
occurrence
of metastable configurations, have been the object of several recent theoretical
studies
\cite{fetter,machida,rok,pu}.
In particular one has to distinguish between configurations
where the system is stable against small deformations of the system
(local stability)
and configurations where the system
is in full thermodynamic equilibrium (global stability).
For frequencies smaller than the
critical frequency (\ref{omegacF}) the system, in the presence of a
vortex, is not thermodynamically stable,
but can be locally stable (metastability) \cite{fetter,machida}. For even
smaller frequencies the vortex is  unstable  also locally
and exhibits elementary
excitations with negative energies \cite{rok,dodd},
unless one considers toroidal geometries.
In the 
following we will discuss only configurations in which the system is in full
thermal equilibrium in the  rotating frame.

We will consider a gas confined by
an axially symmetric harmonic  potential of the form
$V_{ext}=m(\omega_z^2z^2 + \omega_{\perp}^2r^2_{\perp}$)/2 with $r^2_{\perp}=
x^2+y^2$.
Actually, in order to achieve thermalization in the rotating frame
the trap should contain an asymmetry in the $xy$-plane.
One however expects that, unless the gas is too cold,
even very small asymmetries are sufficient to ensure fast
thermalization, so that one can safely use symmetric potentials for the
calculation
of the vortical configuration as well as of the relevant thermodynamic functions
and, at the same time,  assume thermal equilibrium in the rotating frame.

In the absence of two-body interactions the critical frequency $\omega_c$
 is easily calculated
and coincides with the radial oscillator frequency $\omega_{\perp}$.
In interacting gases the  value of $\omega_c$
can be calculated,
at zero temperature, by solving the Gross-Pitaevskii equation
for the order parameter \cite{dalfovo}.
In this case equation (\ref{omegacritical}) takes
 the simpler form $\omega_c=(E_{v}-E_0)/N\hbar$
where $E_{v}$ and $E_0$ are the energies of the ground state
configurations  with and without the vortex, respectively,  and
we have considered a vortex
with one quantum of circulation:
$\langle J_z\rangle_{v}=N\hbar$, $\langle J_z\rangle_0=0$.
 Typical values
of $\omega_c$ correpond to a few tens of Hertz.
Analogously, for each value of $\omega$, one can calculate
a critical value $N_0^c$ such that, for $N_0\ge N_0^c$ the vortex
corresponds to a stable configuration in the frame rotating with frequency
$\omega$.
 Actually the Gross-Pitaevskii
equation exhibits a scaling behaviour in the interaction
parameter $N_0a/a_{ho}$, where
 $a$ is the $s$-wave scattering length, which will be always assumed to be
positive, and  $a_{ho}=\sqrt{\hbar/m\omega_{ho}}$
is the oscillator length
relative to the geometrical average
$\omega_{ho} =
(\omega_z\omega_{\perp}^2)^{1/3}=\lambda^{1/3}\omega_{\perp}$  where
$\lambda = \omega_z/\omega_{\perp}$ is the deformation parameter of the trap.
As a consequence of scaling one can always write
the critical number $N_0^c$ in the form
\begin{equation}
N_0^c(\omega){a\over a_{ho}} = 
f_{\lambda}\left({\omega \over \omega_{\perp}}\right)
\label{scaling}
\end{equation}
where $f_{\lambda} $ is a  function that can be calculated by solving
the Gross-Pitaevskii equation for each value of $\lambda$.
When the dimensionless  parameter 
$N_0a/a_{ho}$ is much larger than $1$ (Thomas-Fermi limit) the
solution of the Gross-Pitaevskii
equation gives the following result \cite{pethick}
(see also \cite{baym,sinha}) for the critical
frequency
\begin{equation}
{\omega_c\over \omega_{\perp}}= {5 \lambda^{1/3}a_{ho}^2\over 2 R_{\perp}^2}
\ln { 0.67 R_{\perp}\over \xi } \, ,
\label{omegacritical}
\end{equation}
where
$R_{\perp} =$ $\lambda^{1/3}a_{ho}
\left(15N_0a /a_{ho}\right)^{1/5}$
 is the Thomas-Fermi radius of the condensate in the $xy$-plane,
while
$\xi = a_{ho} \left(15N_0a /a_{ho}\right)^{-1/5}
\label{xi}$
is the healing length, which fixes the size of the vortex core.
The Thomas-Fermi result (\ref{omegacritical}) allows one to
determine in a simple way the function
$f_{\lambda}$ which turns out to scale as $\lambda^{-5/6}$.
In the opposite limit, when the parameter $N_0a/a_{ho}$ is very small, one can
use a
perturbative treatment of the interaction yielding the result
$\omega_c/\omega_{\perp}= 1 -  \lambda^{1/3}(8\pi)^{-1/2}
N_0a/a_{ho}$.
The quasi ideal gas limit
is not however relevant for the available experimental configurations
since $N_0a/a_{ho}$ is always much larger than $1$.
In fig.1 we compare the results for the function $\lambda^{5/6}f_{\lambda}$,
calculated by
solving the Gross-Pitaevskii equation with the prediction of
the Thomas-Fermi approximation for
two different values of the deformation parameter ($\lambda=\sqrt8$ and
$\lambda=1$).
One sees that the Thomas-Fermi limit is rather accurate for $\omega
\le 0.5\omega_{\perp}$.

From an experimental point of view the creation of a vortex by
adiabatic increase of the rotational frequency of the trap up to $\omega_c$
might  not be   the safest procedure at very low temperature.
In fact one expects that in this case the system will exhibit a high
barrier for the nucleation of vortices as
happens in superfluid helium \cite{donnelly}.
Though the nucleation process in dilute gases might behave
differently from the one of a dense supefluid, one nevertheless expects
that a safer
procedure
to generate vortices is first to rotate the gas at higher temperature
(in case, also above the critical temperature for Bose-Einstein condensation)
and then to cool it via evaporation. This  strategy has motivated
the present investigation of the phase diagram of vortices at finite
temperature.

Let us first discuss the behaviour of the critical temperature $T_c$
as a function of the rotational frequency.
We will suppose the system to be in thermal equilibrium in the frame
rotating with frequency $\omega$.
A first estimate of the critical temperature is obtained by
taking
the ideal gas model. In the
absence of rotations and for large values of $N$ this  model predicts the result
$T_c^0 = 0.94(\hbar\omega_{ho}/k_B)N^{1/3}$. 

The effect of the rotation is to change the shape of the distribution
function of the thermal atoms so that the number of atoms out of the
condensate
takes the form
\begin{eqnarray}
 N_T  & = & {1\over (2\pi \hbar)^3} \int d{\bf r}d{\bf p}
\{ \exp[{p^2\over 2m}  +   {m \over 2}(\omega^2_z z^2 +
\omega_{\perp}^2r^2_{\perp})
\nonumber \\
& - & 
\omega (xp_y-yp_x)]/k_BT -1\}^{-1} \, .
\label{NT}
\end{eqnarray}
Result (\ref{NT}) holds for $T\le T_c$ where the chemical potential of the
non interacting model vanishes in the thermodynamic limit.
Notice that, unless a vortex is formed, the rotation has instead no effect
on the condensate, whose wave function has zero
angular momentum since the confining trap is axially symmetric.
By using the  transformation ${\bf p} \to {\bf p} - m{\bf \omega}\times
{\bf r}$
the  integral (\ref{NT})
takes the same form as in the absence of rotation, with
an effective radial frequency
$\omega_{\perp}(\omega) = (\omega^2_{\perp} - \omega^2)^{1/2}$.
This reflects the role of the centrifugal force which is consequently
responsible for a   lowering of
 the critical temperature $T_c$ according to
\begin{equation}
T_c(\omega) = T_c^0\left(1-{\omega^2 \over \omega_{\perp}^2}\right)^{1/3}
\label{Tc}
\end{equation}
where $T_c^0$
is the critical temperature
in the absence of rotation.
In the same model the
temperature dependence
of the condensate fraction is given by:
\begin{equation}
{N_0\over N} =1-\left({T\over T_c(\omega)}\right)^3 \, .
\label{N0}
\end{equation}
The lowering of $T_c$
is dramatic near  $\omega_{\perp}$ where the
system becomes
 unstable. In this case the centrifugal force exactly cancels the confining
effect of the harmonic
potential. For confining potentials growing faster than $r^2$ at large distances
 the
critical temperature would  instead differ from zero
for any value of the rotating
frequency. The behaviour exhibited by dilute gases in harmonic traps  should
be compared with the one of  dense liquids, like superfluid helium,
where the critical temperature is practically unaffected by the rotation.

The next step is to assume that 
the $T=0$ relationship $N^c_0(\omega)$ of eq.(\ref{scaling}) between 
the frequency $\omega$ and the minimum number of atoms in the condensate
required to have a stable vortex, is not changed at finite
temperature. This relationship allows one to calculate, through
eq.(\ref{N0}), the critical temperature $T_{v}$, below which the vortex
corresponds to a stable configuration in the frame rotating with frequency
$\omega$. We find the useful result
\begin{equation}
T_{v}(\omega) =
T_c^0\left(1-{\omega^2\over\omega^2_{\perp}}\right)^{1/3}
\left(1-{a_{ho} \over a N}f_{\lambda}({\omega \over
\omega_{\perp}})\right)^{1/3}
\label{main}
\end{equation}
where we have explicitly used eq.(\ref{Tc}) for the critical temperature.
 Equation (\ref{main}) points out the crucial role
played by two-body interactions.
In particular,
by increasing the value of the dimensionless parameter $Na/a_{ho}$,
the curve $T_{v}$ is pushed to the left, making the region of stability
for vortices wider and wider.
In fig.2 we show the critical curves $T_c(\omega)$ (dashed line) 
and $T_{v}(\omega)$ (full line).
The function $f_{\lambda}$ entering eq.(\ref{main})
was obtained by solving the Gross-Pitaevskii equation. For temperatures below
  $T_c(\omega)$ the
gas  exhibits Bose-Einstein condensation. However only for temperatures below
$T_{v}(\omega)$ will the vortex  correspond to the equilibrium
configuration of the system.

In deriving the main result (\ref{main})
we have made several approximations that it is worth
discussing explicitly. These concern the use of the ideal gas predictions
(\ref{Tc}) and (\ref{N0}) for the critical temperature and the thermal
depletion of the condensate, as well as the use, at finite
temperature, of the $T=0$ relationship  between the
rotational frequency and the critical
value of $N_0^c$, given by the solution of the
Gross-Pitaevskii equation.

Neglecting interaction effects in the evaluation of the critical
temperature is a very good
approximation since near $T_c$ the gas is extremely dilute. Indeed
self-consistent mean field \cite{jltp,rmp} as well as {\it ab initio}
\cite{krauth} calculations
show that $T_c$ is lowered by interactions only by  a few percent.
The effects on the thermal depletion of the condensate are
instead expected to be more important because of the sizable renormalization
of the chemical potential due to the presence of the condensate.
For the same reason one expects that the relationship
(\ref{scaling}) between the  frequency and the critical  value of  $N_0^c$
 should exhibit a 
temperature dependence.

In order to understand the importance of the above effects let us calculate
the chemical potential $\mu=\partial F/\partial N$ of the gas with and
without the vortex. Also at finite temperature
the chemical potential can be calculated, with
good approximation, by solving the Gross-Pitaevskii equation for the condensate
with a fixed value of $N_0$. In the absence of vortices
and for large values of $N_0a/a_{ho}$ one can use  the Thomas-Fermi formula
$\mu_0(N_0)={1\over 2} \hbar\omega_{ho}(15N_0a/ a_{ho})^{2/5}$.
If a vortex is present
 the chemical potential will contain  an extra
term $\delta\mu = {3\over 5}\hbar \omega_c$ with $\omega_c$ given by
eq.(\ref{omegacritical}).
Notice that both $\mu_0$ and $\delta\mu$ are fixed by  the value
of $N_0$. However, for a given value of $N$ and $T$, the value of $N_0$
differs depending on whether  the vortex is present or not. This effect is
absent
if one uses
 the ideal gas prediction (\ref{N0}). By treating the interaction effect
as a small perturbation one finds the result \cite{rmp}
\begin{equation}
{N_0\over N} = 1 - {T^3\over T_c^3(\omega)} -
{\zeta(2)\over \zeta(3)}{\mu T^2 \over k_BT_c^3(\omega)} \, .
\label{N0+int}
\end{equation}
which generalizes the ideal gas prediction (\ref{N0}).
The above equation shows that, because of interactions, the number of atoms
in the condensate with and without the vortex differ, to the lowest order
in the interaction, by the amount
\begin{equation}
{\delta N_0\over N} =  -
{\zeta(2)\over \zeta(3)}{\delta \mu T^2 \over k_BT_c^3(\omega)} \, .
\label{deltaN0}
\end{equation}
The discontinuity 
$\delta N_0$ in the number of atoms in the condensate is not very
large (at the maximum of the curve $T_{v}$ of fig.2 one finds
$\delta N_0 \sim 200$), but should  be
 taken into account for a safe estimate of the critical frequency.
In fact the
 relevant change in the chemical potential is
given by
$\mu_{v}(N,T) - \mu_0(N,T) = (\partial \mu_0/ \partial N_0) \delta N_0
+ \delta\mu$. By using
(\ref{N0+int}) and (\ref{deltaN0})  one finds
that the   leading corrections, due to the interaction term in (\ref{N0+int}),
cancel out. This suggests that,  at finite temperature, the critical frequency
can be safely calculated using the $T=0$ solution of the Gross-Pitaevskii
equation and the ideal gas expression (\ref{N0}) for $N_0$.
In conclusion  the main result (\ref{main}) for the critical temperature
for vortices is expected to represent  a rather good estimate
also  when  $T_{v}$ is comparable to $T_c$.
 Notice that the
value of $N_0$ on the line $T_{v}$ can be  nevertheless significantly
smaller than the 
ideal gas prediction (\ref{N0}) because of the interaction term
in (\ref{N0+int}). This effect should lower the height
of the barrier at the transition and hence favour the nucleation of the
vortex.

A possible way to approach to the vortical region is through an
adiabatic transformation.
Suppose that, starting from a system initially at rest in the laboratory,
we switch on the rotation of the trap in an adiabatic way. Since the entropy
of the gas is given, in first approximation, by the ideal gas value
$S  = 4k_BN(\zeta(4) / \zeta(3))(T / T_c(\omega))^3$
the adiabatic increase of the frequency has the effect of lowering
the temperature. If the initial temperature
is below $T_c$ the transformation will eventually bring the system into the
vortical region. Notice that this scenario does not require that the initial
temperature
be much smaller than $T_c$.
This is important  for at least two reasons: first
the presence of a large fraction of thermal atoms is crucial to favour the
thermalization of the gas in the rotating frame; second, the nucleation of the
vortex
may become easier if the number of
atoms in the condensate is not too large.

Once the vortex is created a major problem
is its experimental detection. Recently, various methods have been
proposed to detect vortices in these trapped gases. They include the imaging
of the vortex core following the expansion of the atomic cloud
\cite{lund}, the occurrence
of frequency shifts in the collective oscillations
\cite{sinha,zambelli} and the occurrence of
dislocations in the interference patterns between condensates  containing
vortical configurations \cite{bulda}.

In the most interesting vortical region   the system is expected to
 exhibit further interesting  features. In fact
 increasing
the rotational frequency will favour the creation of more complex
vortical configurations, associated with the occurrence of $2$ or more vortices
\cite{rok2,castin}.
In principle vortices with higher quanta of circulation are also possible.
However these configurations are likely unstable \cite{donnelly,pu}.
A final interesting question concerns
the stability of elementary excitations in the rotating gas. In \cite{muntsa}
it has been shown that, for large frequencies, the surface excitations
of the condensate become unstable.
This happens for frequencies larger, but not tremendously larger,
than the critical frequency needed to
create a vortex. For example with $N=10^4$  atoms in
a symmetric trap at $T=0$  it is found \cite{muntsa}
that for
the critical frequency for generating a surface instability is
$\omega/\omega_{\perp} \sim 0.5$, while the one for the vortex instability is
$\sim 0.35$.
 The influence
of the vortex on the stability
of such excitations \cite{machida} as well as the possible occurrence
of new scenarios associated with the "condensation" of surface excitations
are challenging questions for future investigation.

I would like to thank F. Dalfovo, J. Dalibard, S. Giorgini and L. Pitaevskii
for many useful discussions. This project was  supported by
the Istituto Nazionale per la Fisica della Materia through the Advanced
Research Project on BEC.

Figure Captions.

Fig.1. Function $\lambda^{5/6}f_{\lambda}$ calculated
for two different values of $\lambda$
(open circles: $\lambda=1$; solid circles: $\lambda=\sqrt8$). The prediction 
obtained from eq.(\ref{omegacritical}) 
is also reported (dashed line). Frequencies
are in units of $\omega_{\perp}$.

Fig.2. Phase diagram for vortices in a harmonically trapped  Bose gas
($N=10^4$,   $a/a_{ho}= 7.36 \times 10^{-3}$ and $\lambda = 1$). 
Frequencies are in units of $\omega_{\perp}$.

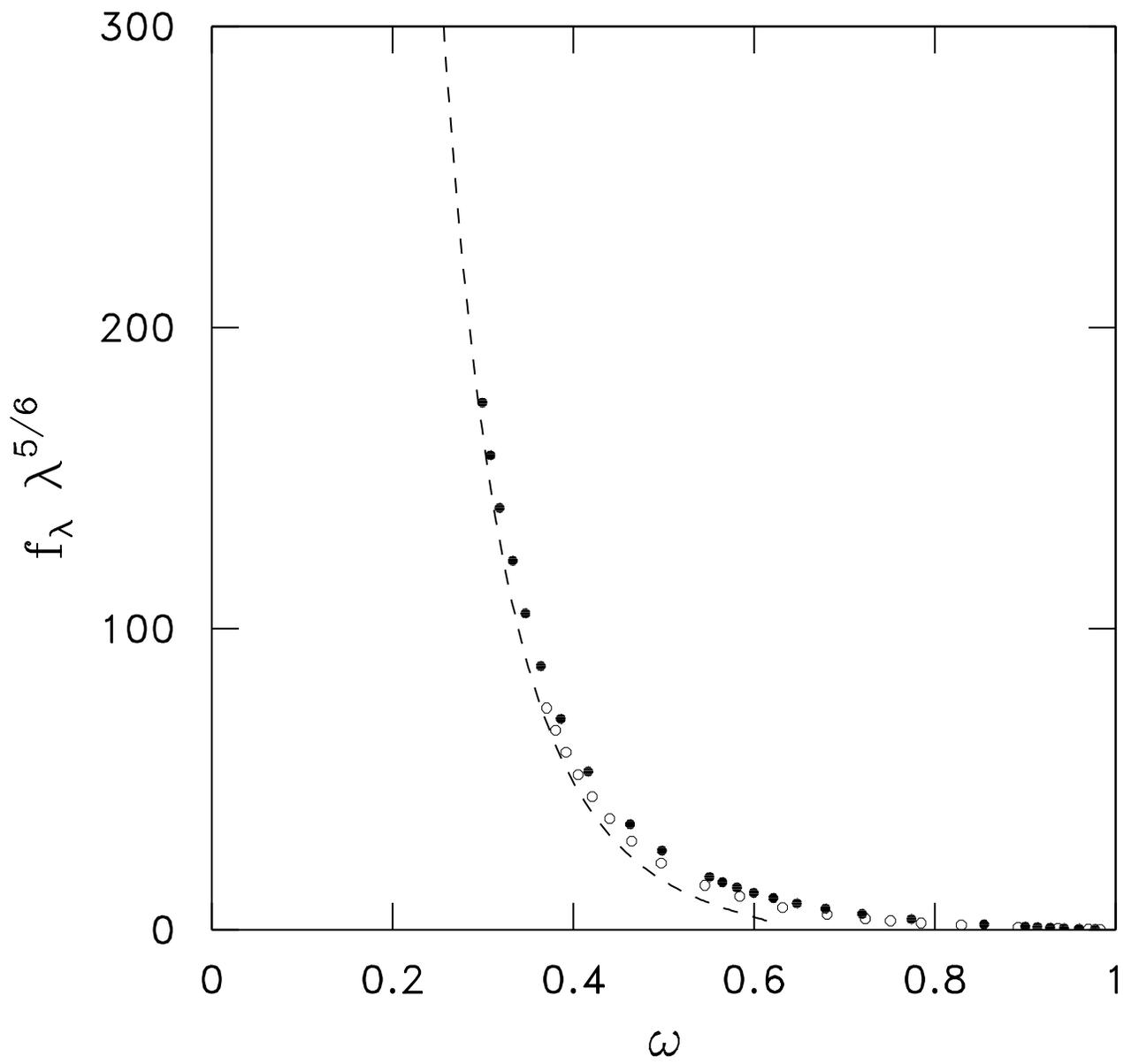
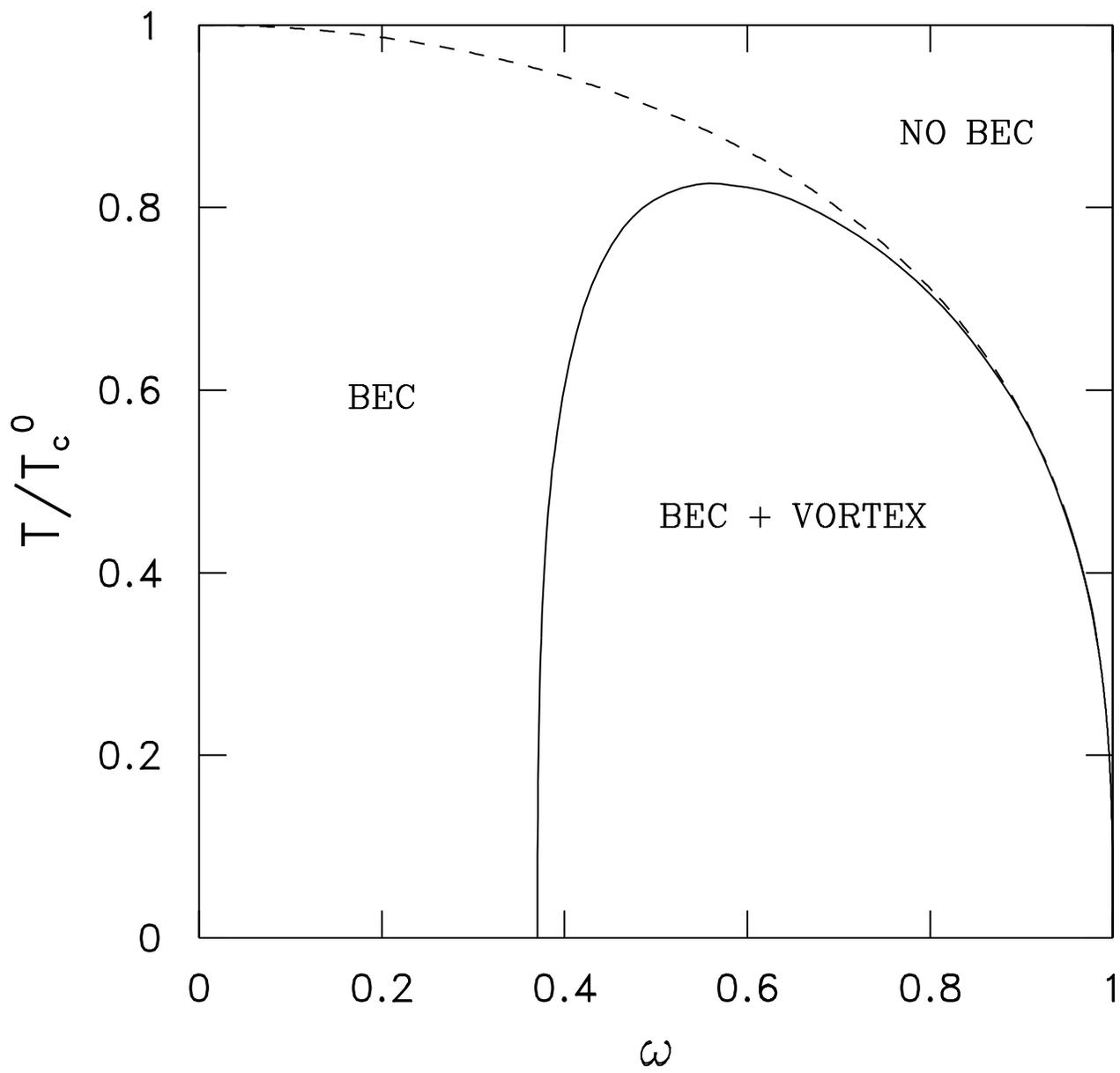
\end{document}